\documentstyle[aps,prb,epsfig,float,twocolumn]{revtex}
\begin{document}
\twocolumn[\hsize\textwidth\columnwidth\hsize\csname
@twocolumnfalse\endcsname

\title{Spin-Hall transport of heavy holes in III-V semiconductor quantum wells}

\author{John Schliemann and Daniel Loss}

\address{Department of Physics and Astronomy, University of Basel,
CH-4056 Basel, Switzerland}

\date{\today}

\maketitle

\begin{abstract}
We investigate spin transport of heavy holes in III-V semiconductor 
quantum wells in the presence of spin-orbit coupling of the Rashba type due to
structure-inversion asymmetry. Similarly to the case of electrons, the
longitudinal spin conductivity vanishes, whereas the off-diagonal elements
of the spin-conductivity tensor are finite giving rise to an intrinsic
spin-Hall effect. For a clean system we find a closed expression
for the spin-Hall conductivity depending on the length scale of the Rashba
coupling and the hole density. In this limit the spin-Hall conductivity
is enhanced compared to its value for electron systems, and it vanishes 
with increasing strength of the impurity scattering.
As an aside, we also derive explicit expressions for the Fermi momenta
and the densities of holes in the different dispersion branches as a function
of the spin-orbit coupling parameter and the total hole density. These results
are of relevance for the interpretation of possible
Shubnikov-de Haas measurements detecting the Rashba spin splitting.
\end{abstract}
\vskip2pc]

\section{Introduction}

In the last years, the emerging field of spin electronics
has become a major branch of solid state physics and comprises by
now all kinds of spin-dependent phenomena in semiconductor
structures and related systems \cite{Wolf01,Awschalom02,Rashba04a}. 
Most recently, the possibility of spin-Hall currents has attracted
considerable theoretical interest
\cite{Murakami03,Culcer03,Hu03,Bernevig03,Hu04,Murakami04,Sinova03,Schliemann03a,Sinitsyn03,Shen03,Burkov03,Rashba03,Inoue04,Shen04,Rashba04b,Dimitrova04}.
In these studies a 
spin current (as opposed to a charge current) driven by an electric field 
perpendicular to it was investigated, where the spinful intinerant
charge carrriers are bulk valence-band holes in III-V 
zinc-blende semiconductors 
\cite{Murakami03,Culcer03,Hu03,Bernevig03,Hu04,Murakami04}
or conduction band electrons in quantum wells 
\cite{Sinova03,Schliemann03a,Sinitsyn03,Shen03,Burkov03,Rashba03,Inoue04,Shen04,Rashba04b,Dimitrova04}
of the same type of materials. In the present work we extend these studies
to the case of heavy holes in quantum wells being subect to spin-orbit 
coupling of the Rashba type due to structure-inversion asymmetry
\cite{Gerchikov92,Winkler00a,Winkler02,Pala02,Rashba88}.

From a historical perspective,
the notion of the spin-Hall effect in systems of itinerant spinful charge
carriers was considered first by Dyakonov and Perel \cite{Dyakonov71} 
in the early seventies, and in a more recent paper by Hirsch \cite{Hirsch99}.
In these studies the predicted spin-Hall effect is due to spin-orbit
effects influencing scattering processes upon static impurities.
Following the terminology used in \cite{Sinova03,Culcer03} this is referred
to as the {\em extrinsic} spin-Hall effect since it necessarily requires 
spin-dependent impurity scattering. This is in contrast to the {\em intrinsic}
spin-Hall effect studied theoretically very recently in 
\cite{Murakami03,Culcer03,Hu03,Bernevig03,Hu04,Murakami04,Sinova03,Schliemann03a,Sinitsyn03,Shen03,Burkov03,Rashba03,Inoue04,Shen04,Rashba04b,Dimitrova04}.
which is entirely
due to spin-orbit coupling terms in the single-particle carrier Hamiltonian
and occurs even in the absence of any scattering process. We note that this 
distinction between intrinsic and extrinsic effects becomes ambiguous in
the limit of weak spin-orbit coupling when life time effects 
of carrier quasiparticles have to be taken into account \cite{Schliemann03a}.

Yet another type of spin-Hall effect was studied recently by Meier and Loss
\cite{Meier03}
in a two-dimensional Heisenberg model consisting of non-itinerant spins,
in contrast to the itinerant-carrier systems mentioned before.

In the case of conduction-band electrons in III-V semiconductor 
quantum wells, the intrinsic spin-Hall effect results from spin-orbit
coupling of the Rashba type (due to structure-inversion asymmetry 
\cite{Rashba60}) and/or the Dresselhaus type (due to bulk-inversion asymmetry 
\cite{Dresselhaus55}). Moreover, the interplay of these two effects leads
to particularly interesting transport effects 
\cite{Schliemann03b,Schliemann03c,Sinitsyn03}. For bulk valence band holes
the effects of spin orbit coupling underlying the 
intrinsic spin-Hall effect are
incorporated in Luttinger's effective Hamiltonian \cite{Luttinger56}
giving rise to two different dispersion branches differing in their
effective masses and known as heavy and light holes. Within this description, 
these states form for a given wave vector multiplets of total angular
momentum 3/2 resulting from the $s=1/2$ spin degree of freedom of the holes and
the $l=1$ orbital anglar momentum of the $p$-type atomic orbitals forming
the valence band \cite{Winkler04a}. 
In a quantum well, a splitting between the heavy and light 
holes occurs due to the size quantization in the growth direction of the 
heterostructure. For a sufficiently narrow quantum well and for not too high
densities and temperatures, only the lowest heavy hole subbands are 
significantly occupied. Here the angular momentum of the heavy hole
states points predominantly along the growth direction \cite{Winkler00a},
corresponding to the total angular momentum states $\pm 3/2$, an approximation
we shall adopt in the present work.

Experimental investigations of
such systems include studies of spin polarization and transitions to an
insulating state induced by magnetic fields \cite{Tutuc01}. Moreover, the
spin splitting due to spin-orbit coupling has been studied in detail
via Shubnikov-de Haas oscillations \cite{Papadakis99,Papadakis02}
including also anisotropic properties of the magnetoresistance
\cite{Papadakis00}. In recent theoretical investigations the
anisotropies in the effective $g$-factor\cite{Winkler00b}, 
spin polarizations\cite{Winkler04b},
and controlled spin rotations induced by Rashba-type spin-orbit
coupling in a spin-FET setup \cite{Pala02} have been considered.
In the present work we investigate the spin-Hall effect for
heavy holes in III-V zinc-blende semiconductor quantum wells.

This paper is organized as follows: In section \ref{model} we discuss the
Rashba model for heavy holes in III-V semiconductor quantum wells. As an
aside, we also give explicit expressions for the Fermi momenta
and the densities of holes in the different dispersion branches as a function
of the spin-orbit coupling parameter and the total hole density. 
Apart from the present investigatios, we expect these results to be also of
relevance for the interpretation of possible Shubnikov-de Haas measurements
\cite{Luo88,Engels97} in such hole systems. 
Section \ref{spinhall} is devoted to the discussion of
our results on spin-Hall transport, and we close with conclusions in
section \ref{conclusions}.

\section{Rashba spin-orbit coupling for heavy holes}
\label{model}

We consider the following single-particle Hamiltonian incorporating
spin-orbit coupling due to structure-inversion asymmetry for heavy holes
in III-V semiconductor quantum wells of appropriate growth geometry
\cite{Gerchikov92,Winkler00a,Winkler02,Pala02,Rashba88},
\begin{equation}
{\cal H}=\frac{\vec p^{2}}{2m}+i\frac{\alpha}{2\hbar^{3}}
\left(p_{-}^{3}\sigma_{+}-p_{+}^{3}\sigma_{-}\right)\,,
\label{defham}
\end{equation}
using the notations $p_{\pm}=p_{x}\pm ip_{y}$, 
$\sigma_{\pm}=\sigma_{x}\pm i\sigma_{y}$, where $\vec p$, $\vec\sigma$
denote the hole momentum operator and Pauli matrices, respectively. 
These Pauli matrices operate on the total angular momentum states
with spin projection $\pm 3/2$ along the growth direction; in this sense they
represent a pseudospin degree of freedom rather than a genuine
spin 1/2. In the above equation, $m$
is the heavy-hole mass, and $\alpha$ is Rashba spin-orbit coupling coefficient 
due to structure inversion asymmetry across the quantum well grown along the
[001]-direction chosen as the
$z$-axis. For a symmtrically grown quantum well, the coefficient $\alpha$ is 
essentially proportional to an electric field applied across the well and
therefore experimentally tunable \cite{Gerchikov92,Winkler00a,Winkler02}.
This Hamiltonian has two dispersion branches given by
\begin{equation}
\varepsilon_{\pm}(k)=\frac{\hbar^{2}k^{2}}{2m}\pm\alpha k^{3}
\label{dispersion}
\end{equation}
with eigenfunctions
\begin{equation}
\langle\vec r|\vec k,\pm\rangle=
\frac{e^{i\vec k\vec r}}{\sqrt{A}}\frac{1}{\sqrt{2}}
\left(
\begin{array}{c}
1 \\
\mp i\left(k_{x}+ik_{y}\right)^{3}/k^{3}
\end{array}
\right)\,,
\end{equation}
where $\vec k$ is the hole lattice momentum and $A$ is the area of the system.
We note that the validity of above model given by the Hamiltonian 
(\ref{defham}) is restricted to sufficiently small wave numbers and densities.
In fact, the lower of the two dispersion branches (\ref{dispersion}) is 
(for $\alpha>0$) not 
bounded from below for large wave numbers and has non-negative eigenenergies
only for $k \leq\hbar^{2}/2m\alpha$ with a maximum at 
$k=\hbar^{2}/3m\alpha$. 
The unbounded decrease of the single-particle energies with increasing
wave number for $k>\hbar^{2}/3m\alpha$ is clearly an unphysical feature of the 
model. Therefore, the following 
considerations are restricted to sufficiently small
densities such that (at zero temperature) only states with 
wave numbers $k \leq\hbar^{2}/3m\alpha$ are occupied. 
  
\begin{figure}
\centerline{\includegraphics[width=8cm]{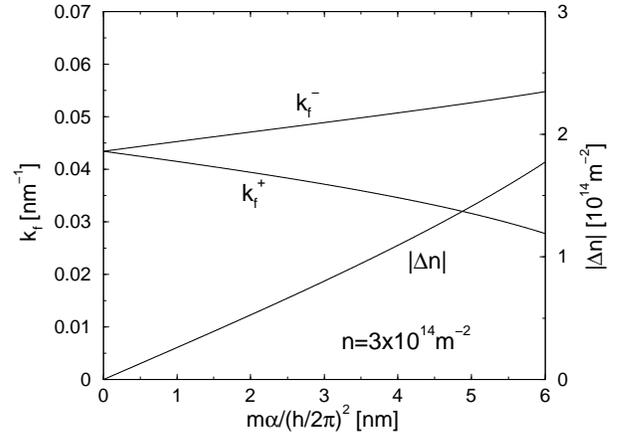}} 
\caption{The Fermi wave numbers $k_{f}^{\pm}$ and the difference $\Delta n$
of hole densities in the two dispersion branches as a function of the
characteristic length $m\alpha/\hbar^{2}$ of the Rashba coupling at a total
hole density of $n=3\cdot 10^{14}{\rm m^{-2}}$. 
\label{fig1}}
\end{figure}

For a given Fermi energy $\varepsilon_{f}$
and vanishing temperature, the above two dispersion branches
give rise to two different Fermi wave numbers $k_{f}^{\pm}$ fulfilling
\begin{equation}
\varepsilon_{f}=\frac{\hbar^{2}\left(k_{f}^{\pm}\right)^{2}}{2m}
\pm\alpha\left(k_{f}^{\pm}\right)^{3}
\label{defk}
\end{equation}
with $k_{f}^{+}<k_{f}^{-}$ ($k_{f}^{+}>k_{f}^{-}$) for positive (negative)
$\alpha$. Subtracting these two equations one finds
\begin{equation}
0=\frac{\hbar^{2}}{2m}
\left(\left(k_{f}^{+}\right)^{2}-\left(k_{f}^{-}\right)^{2}\right)
+\alpha\left(\left(k_{f}^{+}\right)^{3}+\left(k_{f}^{-}\right)^{3}\right)
\label{keq1}
\end{equation}
or, cancelling a factor of $(k_{f}^{+}+k_{f}^{-})$,
\begin{equation}
k_{f}^{+}k_{f}^{-}=-\frac{\hbar^{2}}{2m\alpha}
\left(k_{f}^{+}-k_{f}^{-}\right)-\left(k_{f}^{+}-k_{f}^{-}\right)^{2}\,.
\label{keq2}
\end{equation}
Note that in the relations (\ref{keq1}), (\ref{keq2}) the Fermi energy does not
enter explicitly. In fact, the quantity which can be immediately controlled
experimentally is not the Fermi energy but the hole density $n$ given by
\begin{equation}
n=\frac{1}{4\pi}
\left(\left(k_{f}^{+}\right)^{2}+\left(k_{f}^{-}\right)^{2}\right)
\label{density}
\end{equation}
Combining Eqs.~(\ref{keq2}), (\ref{density}), one obtains
\begin{eqnarray}
k_{f}^{+}-k_{f}^{-} & = & -\frac{\hbar^{2}}{2m\alpha}
\left(1-\sqrt{1-\left(\frac{2m\alpha}{\hbar^{2}}\right)^{2}4\pi n}\right)\,,
\label{kdiff}\\
k_{f}^{+}k_{f}^{-} & = & 4\pi n\nonumber\\
 & - & \left(\frac{\hbar^{2}}{2m\alpha}\right)^{2}
\left(1-\sqrt{1-\left(\frac{2m\alpha}{\hbar^{2}}\right)^{2}4\pi n}\right)\,.
\label{kprod}
\end{eqnarray}
Note that for $k_{f}^{\pm} \leq\hbar^{2}/3m\alpha$ the radicand in the
above equations is always positive. Using 
\begin{equation}
4\pi n =\left(k_{f}^{+}+k_{f}^{-}\right)^{2}-2k_{f}^{+}k_{f}^{-}
\end{equation}
one derives from the above equations the following explicit expression
for $k_{f}^{\pm}$ as a function of the density $n$ and the Rashba parameter
$\alpha$:
\begin{eqnarray}
& & k_{f}^{\pm}=\mp\frac{1}{2}\frac{\hbar^{2}}{2m\alpha}
\left(1-\sqrt{1-\left(\frac{2m\alpha}{\hbar^{2}}\right)^{2}4\pi n}\right)
\nonumber\\
& & \quad+\Biggl[
-\frac{1}{2}\left(\frac{\hbar^{2}}{2m\alpha}\right)^{2}
\left(1-\sqrt{1-\left(\frac{2m\alpha}{\hbar^{2}}\right)^{2}4\pi n}\right)
\nonumber\\
 & & \quad\quad+3\pi n\Biggr]^{1/2}\,.
\label{kfpm}
\end{eqnarray}
The difference $\Delta n$ of densities of holes in the two dispersion branches,
\begin{equation}
\Delta n=\frac{1}{4\pi}
\left(\left(k_{f}^{+}\right)^{2}-\left(k_{f}^{-}\right)^{2}\right)
\label{diffdensity}\,,
\end{equation}
can be expressed as
\begin{eqnarray}
& & \Delta n=\frac{-1}{4\pi}\frac{\hbar^{2}}{2m\alpha}
\left(1-\sqrt{1-\left(\frac{2m\alpha}{\hbar^{2}}\right)^{2}4\pi n}\right)
\nonumber\\
& & \quad\cdot\Biggl[
-2\left(\frac{\hbar^{2}}{2m\alpha}\right)^{2}
\left(1-\sqrt{1-\left(\frac{2m\alpha}{\hbar^{2}}\right)^{2}4\pi n}\right)
\nonumber\\
 & & \quad\quad+12\pi n\Biggr]^{1/2}\,.
\end{eqnarray}
Such a difference in densities in the two dispersion branches could 
be experimentally detected in Shubnikov-de Haas measurements, as it was shown
earlier for the case of electrons \cite{Luo88,Engels97}. We note that the
above results for the Fermi momenta and $\Delta n$ depend only on the
total hole densty $n$ and length scale $m\alpha/\hbar^{2}$ given by the
Rashba coupling \cite{note1}. Winkler {\em et al.} \cite{Winkler02} have
studied both theoretically and experimentally 
the magnitude of the Rashba spin orbit coupling in GaAs-based
quantum well samples with heavy-hole densities of a few $10^{14}{\rm m^{-2}}$
and have found typical values for the characteristic length scale
$m\alpha/\hbar^{2}$ of a few nanometers. In figure~\ref{fig1} we have 
plotted the Fermi wave numbers $k_{f}^{\pm}$ and the difference $\Delta n$
of holes in the two dispersion branches as a function of $m\alpha/\hbar^{2}$ 
at a total hole density of $n=3\cdot 10^{14}{\rm m^{-2}}$.
 
On the other hand, solving Eq.~(\ref{kdiff}) for $\alpha$ gives a convenient
expression for the Rashba coefficient as a function of the total hole
density and the difference of the Fermi wave numbers,
\begin{equation}
\alpha=\frac{\hbar^{2}}{2m}\frac{-2\left(k_{f}^{+}-k_{f}^{-}\right)}
{\left(k_{f}^{+}-k_{f}^{-}\right)^{2}+4\pi n}\,.
\end{equation} 

Finally, writing Eq.~(\ref{defk}) in the form
\begin{equation}
\varepsilon_{f}\left(\frac{1}{k_{f}^{\pm}}\right)^{3}=\frac{\hbar^{2}}{2m}
\frac{1}{k_{f}^{\pm}}\pm\alpha
\label{fermik1}
\end{equation}
one obtains by adding these two equations
\begin{equation}
\frac{2m}{\hbar^{2}}\varepsilon_{f}
\left(\left(\frac{1}{k_{f}^{+}}-\frac{1}{k_{f}^{-}}\right)^{2}
+\frac{1}{k_{f}^{+}k_{f}^{-}}\right)=1\,,
\label{fermik2}
\end{equation}
which does not explicitly depend on the Rashba parameter $\alpha$.
From this relation, it follows with the help of Eq.~(\ref{keq2})
\begin{equation}
\varepsilon_{f}=\alpha\frac{\left(k_{f}^{+}k_{f}^{-}\right)^{2}}
{k_{f}^{-}-k_{f}^{+}}\,.
\end{equation}
Using Eqs.~(\ref{kdiff}), (\ref{kprod}) it is straightforward to obtain from
this expression an explicit relation between the Fermi energy $\varepsilon_{f}$
and the hole density $n$ at a given Rashba parameter $\alpha$.

\section{Spin-Hall transport}
\label{spinhall}

We now investigate the possibilty of spin-Hall transport of heavy holes
in III-V semiconductor quantum wells in the presence of Rashba-type spin-orbit
coupling. We are interested in spin currents (as opposed to charge currents)
as the linear response of the system to an electric field applied in the 
plane of the well. As we shall see below, in such a system the
 spin current is always perpendicular to the driving electric field and 
therefore Hall-type. We concentrate on the case of zero temperature.

The single-particle spin current operator is defined by
\begin{equation}
\vec j^{S,z}=\frac{3\hbar}{2}\frac{1}{2}
\left(\sigma^{z}\vec v+\vec v\sigma^{z}\right)
=\frac{3\hbar}{2}\frac{\vec p}{m}\sigma^{z}\,,
\end{equation}
where the factor of $3/2$ reflects the angular momentum quantum numbers of the
heavy holes. The hole velocity operator reads 
$\vec v=i[{\cal H},\vec r]/\hbar$ with $\vec r$ being the position operator,
or, in terms of components, 
\begin{eqnarray}
v_{x} & = & \frac{p_{x}}{m}
+\frac{\alpha}{\hbar^{3}}\left(6p_{x}p_{y}\sigma^{x}
+3\left(p_{y}^{2}-p_{x}^{2}\right)\sigma^{y}\right)\,,\\
v_{y} & = & \frac{p_{y}}{m}
+\frac{\alpha}{\hbar^{3}}\left(
3\left(p_{x}^{2}-p_{y}^{2}\right)\sigma^{x}
+6p_{x}p_{y}\sigma^{y}\right)\,.
\end{eqnarray}
A rigorous expression for the spin conductivity, i.e. the linear transport
coefficient for spin currents driven by a spatially homogeneous 
in-plane electric field,
is given by the Kubo formula with full frequency dependence for an
electric field \cite{Mahan00},
\begin{eqnarray}
\sigma^{S,z}_{xy}(\omega) & = & \frac{e}{A(\omega+i\eta)}
\int_{0}^{\infty}e^{i(\omega+i\eta)t}\nonumber\\
 & & \cdot\sum_{\vec k,\mu}f(\varepsilon_{\mu}(\vec k))
\langle\vec k,\mu|[j^{S,z}_{x}(t),v_{y}(0)]|\vec k,\mu\rangle\,,
\label{generalKubo}
\end{eqnarray}
where we have concentrated on the off-diagonal $xy$-components. Moreover,
we have assumed zero temperature $T=0$ and non-interacting carriers,
which allows to formulate the two-body Green's function entering the
conductivity Kubo formula in terms of single-particle operators.
$e=|e|$ is the elementary charge, and
$f(\varepsilon_{\mu}(\vec k))$ is the $T=0$ Fermi distribution function 
for  energy $\varepsilon_{\mu}(\vec k)$ at wave vector $\vec k$ in the
dispersion branch $\mu\in\{+,-\}$.
The spin-current operator (in the Dirac picture)
for spin moment polarized along the $z$-direction and flowing in the
$x$-direction is given by 
\begin{equation}
j^{S,z}_{x}(t)=
e^{i{\cal H}t/\hbar}j^{S,z}_{x}(0)e^{-i{\cal H}t/\hbar}
=\frac{3\hbar}{2m}\sigma^{z}(t)p_{x}(t)\,.
\end{equation}
From now on we will assume the Hamiltonian generating the above
time evolution to include also scattering potentials from
static random impurities being present in the quantum well.
The right hand side of Eq.~(\ref{generalKubo}) is to be understood
in the limit of vanishing imaginary part $\eta>0$ in the frequency argument.
This imaginary part in the frequency reflects the
fact that the external electric field is assumed to be switched on 
adiabatically starting from the infinite past of the system, and it also
ensures causality properties of the retarded Green's function occurring in
Eq.~(\ref{generalKubo}). In general the limiting process 
$\eta\to 0$ does not commute with other
limits, and, in particular, the dc-limit $\omega\to 0$ has to be taken with 
care \cite{Mahan00}. In the presence of random impurity scattering,
the retarded two-body Green's function in Eq.~(\ref{generalKubo}) will 
generically have
a frequency argument with positive imaginary part \cite{Mahan00}. 
In this case the limit 
$\eta\to 0$ is unproblematic, and the imaginary part of the
frequency argument is just due to impurity scattering and/or other
(many-body) effects. For the present problem of impurity scattering of
non-interacting carriers being subject to Rashba-type spin-orbit coupling,
the resulting imaginary part $\eta>0$ in the frequency argument 
is given, to lowest order in the Rashba coefficient and the
impurity potential, by the inverse of the momentum relaxation time.
This is certainly a very intuitive result; the formal arguments leading to it 
is completely analogous to the ones used in Ref.~\cite{Schliemann03a}
and can be given along the following lines: 
In lowest order in the
spin-orbit coupling and the impurity scattering the time-dependent 
spin-current operator reads
\begin{equation}
j^{S,z}_{x}(t)\approx\frac{3\hbar}{2m}\sigma^{z}_{0}(t)p_{x}^{0}(t)
\end{equation}
where the time evolution of $\sigma^{z}_{0}$ is only due to the
Hamiltonian (\ref{defham}), while $p_{x}^{0}(t)$ contains
the impurity scattering but not the spin-orbit coupling.
Now it is useful to note that, in order to compute the
expectation values in the Kubo formula Eq.~(\ref{generalKubo}),
only matrix elements of the time-dependent momentum operator
$p_{x}^{0}(t)$ which are diagonal in the wave vector index are needed. 
This enables to apply superoperator
techniques developed in Refs.~\cite{Loss86} yielding
\begin{equation}
\left(p_{x}^{0}(t)\right)_{\vec k\vec k}\approx
\left(e^{-\Omega_{0}t}p_{x}^{0}(0)\right)_{\vec k\vec k}
=\left(e^{-t/\tau}p_{x}^{0}(0)\right)_{\vec k\vec k}\,,
\label{momentumrelax}
\end{equation}
where $\Omega_{0}$ is the scattering master operator in lowest 
order of the scattering potential \cite{Loss86}. This operator is the same
as it occurs as the scattering term in the usual Boltzmann
equation when evaluated in lowest oder via Fermi's golden rule.
Thus, Eq.~(\ref{momentumrelax}) describes the usual momentum relaxation
due to static impurities in lowest order in the scattering potential.
For impurity potentials being isotropic in real space, the momentum 
$p_{x}$ is an exact eigenfunction of $\Omega_{0}$, and the 
eigenvalue is given by the well-known inverse momentum relaxation 
time $1/\tau(\varepsilon)$ 
\cite{Loss86,Smith89} which in general depends on the energy
$\varepsilon(\vec k)$. To lowest order in the Rashba coupling,
this energy argument can be replaced with the Fermi energy in the absence of
spin-orbit interaction. We note that this momentum relaxation rate
$1/\tau$ is the same as obtained in the standard diagrammatic approach
and thus contains the vertex correction for particle transport \cite{Mahan00}. 
However, this vertex correction vanishes for short-range isotropic scatterers.

The question of vertex corrections to the spin-Hall transport was also
examined very recently by Inoue, Bauer, and Molenkamp~\cite{Inoue04},
and by Dimitrova \cite{Dimitrova04}
for the case of electrons
being subject to Rashba spin-orbit interaction. There the authors reach the
conclusion that for small but finite disorder the spin-Hall conductivity
should identically vanish due to vertex corrections, while it has
its ``universal value'' of $e/8\pi$ in the case of a perfectly 
clean system \cite{Sinova03,Schliemann03a,Sinitsyn03}. Moreover, Murakami
\cite{Murakami04} has studied vertex correction to spin-Hall transport
of holes in bulk p-type semiconductors described by a Luttinger Hamiltonian
\cite{Luttinger56}. There the author concludes that vertex correction vanish 
identically, validating, the results of Ref.~\cite{Schliemann03c}, and 
ascribes this observation to the fact that the underlying Hamiltonian
is invariant under inversion of momenta, in contrast to the Rashba Hamiltonian.
These technically rather involved issues are to the opinion of the present
authors not entirely settled yet and presently under investigation. 

Let us now turn to the evaluation of the spin-Hall conductivity
using the aforementioned  approximations. A straightforward calculation yields
\begin{eqnarray}
\sigma^{S,z}_{xy}(\omega) & = & -\sigma^{S,z}_{yx}(\omega)\nonumber\\
 & = & -\frac{e}{\pi}\frac{9}{4}\frac{\alpha}{m}\int_{k_{f}^{+}}^{k_{f}^{-}}dk
\frac{k^{4}}{\left(\omega+\frac{i}{\tau}\right)^{2}-
\left(\frac{2\alpha k^{3}}{\hbar}\right)^{2}}\,,
\label{fullspinHall}
\end{eqnarray}
where, according to the above arguments, the imaginary part of the
frequency argument is given by the momentum relaxation rate $1/\tau$.
Moreover, the {\em longitudinal} spin conductivities $\sigma^{S,z}_{xx}$,
$\sigma^{S,z}_{yy}$ turn out ot be identically zero. This is similar to the
case of electrons in a quantum well being subject to spin-orbit 
coupling of {\em either} the Rashba or the Dresselhaus type
\cite{Sinova03,Schliemann03a,Sinitsyn03}. There a longitudinal
spin conductivity occurs only if both the Rashba and the Dresselhaus
coupling are present \cite{Sinitsyn03}.

\begin{figure}
\centerline{\includegraphics[width=8cm]{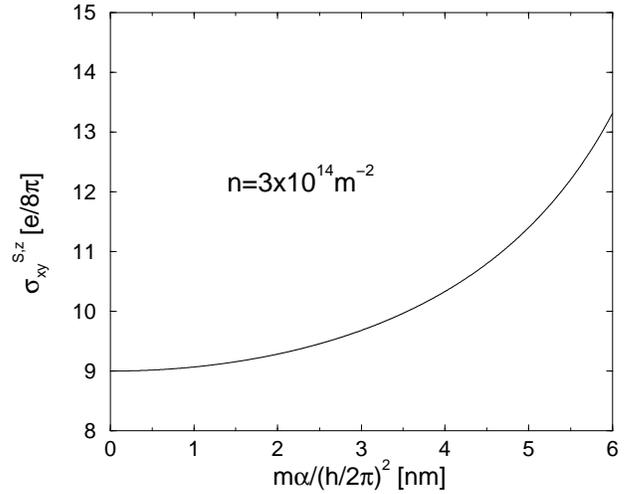}} 
\caption{The spin-Hall condcutivity for dominating Rashba coupling 
($\varepsilon_{R}\gg\hbar/\tau$, cf. Eq.~(\ref{spinHall2})) 
as a function of $m\alpha/\hbar^{2}$
at a total hole density of $n=3\cdot 10^{14}{\rm m^{-2}}$.
\label{fig2}}
\end{figure}

The remaning integral in the above expression (\ref{fullspinHall})
is elementary; however, 
it leads to a rather cumbersome expression which shall not be given here.
In the dc limit $\omega=0$, the energy scale of the impurity scattering
$\hbar/\tau$ has to be compared with the ``Rashba energy'' 
$\varepsilon_{R}=\alpha (k_{f}^{0})^{3}$, where 
$k_{f}^{0}=\sqrt{2m\varepsilon_{f}/\hbar^{2}}$ is the Fermi wave number
for vanishing spin-orbit coupling, which is a typical value
for $k$ in the inegration in Eq.~(\ref{fullspinHall}).
If the impurity scattering dominates over the Rashba coupling,
$\hbar/\tau\gg\varepsilon_{R}$, the spin-Hall conductivity vanishes
with the leading order correction given by
\begin{eqnarray}
\sigma^{S,z}_{xy}(0) & = & \frac{e}{\pi}\frac{9}{20}\frac{\alpha}{m}\tau^{2}
\left((k_{f}^{-})^{5}-(k_{f}^{+})^{5}\right)\nonumber\\
& + & {\cal O}\left(\left(\frac{\varepsilon_{R}}{\hbar/\tau}\right)^{4}\right)
\,,
\label{spinHall1}
\end{eqnarray}
where the Fermi wave numbers are given by Eq.~(\ref{kfpm}). In the opposite 
case $\varepsilon_{R}\gg\hbar/\tau$, the leading contribution to the 
spin-Hall conductivity reads
\begin{eqnarray}
\sigma^{S,z}_{xy}(0) & = &\frac{e}{\pi}\frac{9}{16}
\frac{\hbar^{2}}{m\alpha}
\left(\frac{1}{k_{f}^{+}}-\frac{1}{k_{f}^{-}}\right)\nonumber\\
& + & {\cal O}\left(\left(\frac{\hbar/\tau}{\varepsilon_{R}}\right)^{4}\right)
\,.
\label{spinHall2}
\end{eqnarray}
Note that this result for the spin-Hall conductivity depends only on the
length scale $m\alpha/\hbar^{2}$ of the Rashba coupling and the total hole
density $n$, but not separately on quantities like the Fermi energy and
the effective heavy hole mass. If $m\alpha/\hbar^{2}$ is small against
the inverse square root of the total hole density (but still fulfilling
$\varepsilon_{R}\gg\hbar/\tau$), the spin-Hall conductivity
approaches a value of $\sigma^{S,z}_{xy}=9e/8\pi$. 
This is the case if $\hbar/\tau\ll\varepsilon_{R}\ll\varepsilon_{f}$. 
This above value should be  
compared with the universal value of $e/8\pi$ found for the spin-Hall
conductivity of electrons being subject to a Rashba coupling dominating
possible impurity scattering \cite{Sinova03,Schliemann03a}. 
Thus, in this limit, the hole spin-Hall conductivity is enhanced 
compared to the case of electrons by a factor of $18$, 
which is partially due to the larger
angular momentum of the heavy holes. In figure~\ref{fig2} we have plotted
the spin-Hall conductivity for dominating Rashba coupling 
($\varepsilon_{R}\gg\hbar/\tau$, cf. Eq.~(\ref{spinHall2})) 
as a function of $m\alpha/\hbar^{2}$
at a total hole density of $n=3\cdot 10^{14}{\rm m^{-2}}$. As seen there, the 
spin-Hall conductivity starts out at $\sigma^{S,z}_{xy}=9(e/8\pi)$
and increases with increasing Rashba coupling.

\section{Conclusions}
\label{conclusions}

We have studied spin transport of heavy holes in III-V semiconductor 
quantum wells in the presence of spin-orbit coupling of the Rashba type due to
structure-inversion asymmetry. Similarly to the case of electrons, the
longitudinal spin conductivity vanishes, whereas the off-diagonal elements
of the spin-conductivity tensor are finite giving rise to an intrinsic
spin-Hall effect. For a clean system we find a closed expression
for the spin-Hall conductivity depending on the length scale of the Rashba
coupling and the hole density. In this limit the spin-Hall conductivity
is enhanced compared to its value for electron systems. For dirtier
p-doped quantum wells when the impurity scattering dominates the 
spin-orbit coupling, the spin-Hall conductivity naturally vanishes as
also found previously for the case of electrons 
\cite{Schliemann03a,Sinitsyn03}.
As an aside , we give explicit expressions for the Fermi momenta
and the densities of holes in the different dispersion branches as a function
of the spin-orbit coupling parameter and the total hole density. These results
are expected to be helpful for the interpretation of possible
Shubnikov-de Haas experiments aiming at the detection of the 
Rashba spin splitting.

{\em Note added}: After submission of this paper a preprint by
Wunderlich {\em et al} \cite{Wunderlich04}
appeared reporting on an experimental observation
of the spin Hall effect in p-doped GaAs quantum wells as studied here.

\acknowledgments{We thank J. Carlos Egues, S. Erlingsson, C.-M. Hu, D. Saraga,
O. Shalaev, and R. Winkler for stimulating discussions. 
This work was supported by the NCCR Nanoscience, the Swiss NSF, DARPA, ARO,
ONR, and the EU Spintronics RTN.}



\begin{references}
\bibitem{note1}
We note that for Rashba coupling of conduction band electrons the
characteristic length scale is given by $\hbar^{2}/m\tilde\alpha$ where
$\tilde\alpha$ is the electron Rashba parameter. This quantity is naively
speaking just the ``inverse'' of the characteristic length $m\alpha/\hbar^{2}$
for holes. This difference is due to the different functional form of the 
Rashba spin-orbit contribution for electrons and holes
(linear versus cubic dependence on the momenta). Moreover, the length scale
$m\alpha/\hbar^{2}$ introduced here is different from the spin precession
length used in Ref.~\protect{\cite{Pala02}} which involves the Fermi energy.
\end{references}
\end{document}